\documentclass[letterpaper,twocolumn,superscriptaddress, showpacs,preprintnumbers,amsmath,amssymb] {revtex4}
\usepackage{graphicx}
\usepackage{dcolumn}
\usepackage{bm}
\usepackage{hyperref}
\hypersetup{
    colorlinks,
    citecolor=red,
    filecolor=blue,
    linkcolor=blue,
    urlcolor=blue
}

\newcommand{\Tr}{\mathrm{tr}}
\newcommand{\eq}{\mathrm{eq}}
\newcommand{\MC}{\mathrm{MC}}

\begin{document}

\title{Testing whether all eigenstates obey the Eigenstate Thermalization Hypothesis}

\author{Hyungwon Kim}

\affiliation{Physics Department, Princeton University, Princeton, NJ 08544, USA}

\author{Tatsuhiko N. Ikeda}
\affiliation{Department of Physics, University of Tokyo, Bunkyo-ku, Tokyo 113-0033, Japan}

\author{David A. Huse}
\affiliation{Physics Department, Princeton University, Princeton, NJ 08544, USA}

\begin{abstract}
We ask whether the Eigenstate Thermalization Hypothesis (ETH) is valid
in a strong sense: in the limit of an infinite system, {\it every} eigenstate
is thermal.
We examine expectation values of few-body operators in highly-excited many-body eigenstates
and search for `outliers', the eigenstates that deviate the most from ETH.
We use exact diagonalization of two one-dimensional nonintegrable models:
a quantum Ising chain with transverse and longitudinal fields, and hard-core bosons at half-filling
with nearest- and next-nearest-neighbor hopping and interaction.
We show that even the most extreme outliers appear to obey ETH as the system size increases,
and thus provide numerical evidences that support ETH in this strong sense.
Finally,
periodically driving the Ising Hamiltonian, we show that the eigenstates of the corresponding Floquet operator obey ETH
even more closely.
We attribute this better thermalization to removing the constraint of conservation of the total energy.
\end{abstract}


\maketitle

\section{Introduction}
A cup of hot coffee and a glass of cold beer empirically
thermalize to room temperature if they are left out.
This conventional description of thermalization assumes an infinitely large `reservoir'
with which an initially out-of-equilibrium system can exchange energy (and particles if allowed)
to achieve thermal equilibrium.
However, this does not describe thermalization of an {\it isolated} system.
When a large isolated system thermalizes, what happens is that small subsystems thermalize, with the
remainder of the full system serving as the reservoir.
The question of how this works for a quantum
many-body system undergoing linear unitary time evolution has a long history
(see e.g. \cite{neumann,shnirelman, berry}),
and has attracted renewed attention
due to experimental realizations of such systems~\cite{polkovnikov,yukalov}.

Even though unitary quantum time-evolution is reversible and retains
all information about the initial state,
thermalization can occur if one restricts observations to only few-body operators,
and can not access the full many-body density operator.
This occurs because
all the information in the initial state spreads over the entire system in the long-time limit,
and its reconstruction requires access to the full many-body density operator.

The Eigenstate Thermalization Hypothesis (ETH)
has been proposed as the underlying mechanism
for thermalization in isolated quantum systems
based on random matrix~\cite{deutsch}
and semiclassical~\cite{srednicki} arguments.
ETH states that within the eigenstates of a nonintegrable (quantum chaotic) Hamiltonian,
few-body operators have thermal distributions in the thermodynamic limit,
{\it i.e.} the same probability distributions as the Boltzmann-Gibbs ensemble at the corresponding temperature.

Numerical
tests of the ETH have recently been performed for a wide variety of nonintegrable models \cite{ro,rigol1,santos,rs,khatami1,ikeda,beugeling,steinigeweg,kruczenski,sorg}.
In every case, good numerical evidence is provided that the deviation of a typical eigenstate from thermal value decreases towards zero as the
size of the system is increased, as expected from the ETH.  However, most of these studies only test that
\textit{almost all} energy eigenstates
obey the ETH, while ETH is actually expected to make the stronger claim that
\textit{all} eigenstates away from the edges of the many-body spectrum are thermal in the thermodynamic limit \cite{thermal_states}.
This difference is of fundamental importance, because it is also the difference between
\textit{almost all} initial states approaching thermal equilibrium,
and \textit{all} initial states doing so.
Some previous works have tested whether all eigenstates obey ETH by analyzing
the mismatch between eigenstate expectation values and microcanonical ensemble averages \cite{santos,khatami1}.

In this paper, we provide more numerical evidence that \textit{all} eigenstates obey ETH for
two one-dimensional nonintegrable models:
an Ising chain with transverse and longitudinal fields and hard-core bosons at half-filling
with nearest- and next-nearest-neighbor hopping and interaction.
We also show that driving the Ising chain periodically in time, which removes the conservation of total energy,
makes its eigenstates obey ETH more closely for a given finite-size system.

One remark is in order.
There are two notable exceptions to ETH; integrable systems and many-body localized (MBL) systems.
An integrable system has an extensive number of local conservation laws.
Therefore, the conventional Gibbs ensemble is not enough to fix the probability distributions of
few-body operators.  It is conjectured that distributions of few-body operators
in integrable systems are given by the Generalized Gibbs Ensemble (GGE) \cite{rigol,calabrese, eisert2, fagotti},
however some recent papers report insufficiencies of the GGE in certain cases \cite{goldstein,caux1,caux2,pozsgay,prosen1,ga}.
Many-body localization happens when quenched disorder is strong enough in an interacting system (see Refs. \cite{anderson, baa,nandkishore}
and references therein).
In a MBL phase, individual many-body eigenstates are localized and violate ETH \cite{khatami1}, due to localized conserved observables.
Although integrable systems and MBL systems are exciting and currently very active topics of research,
we restrict ourselves to disorder-free, robustly nonintegrable systems throughout this paper.

The rest of this paper is organized as follows.
In Section II,
we develop an ETH indicator to investigate the strong ETH
and describe the two models investigated in our study.
In Section III,
using exact diagonalization,
we conduct an `outlier' analysis.
The outlier states are the energy eigenstates
that deviate the most from the predictions of ETH.
We show that
even these outlier states converge towards ETH behavior
as we increase the system size.
We also investigate some properties such as the momenta and the participation ratios of the `outlier' states.
In Section IV,
we periodically drive the Ising chain by two sets of non-commuting operators in the Hamiltonian
and study the eigenstates of the corresponding Floquet operator, which are
thermal at infinite temperature \cite{luca,lazarides,ponte}.
Again doing an outlier analysis, we show that the eigenstates of the Floquet operator of a finite chain
deviate from ETH significantly less than do the eigenstates of the corresponding time-independent Hamiltonian
(and of course these deviations vanish in the thermodynamic limit).

\section{ETH indicator and Models}
\subsection{ETH indicator}
Given a time-independent Hamiltonian $H$ and an out-of-equilibrium initial state $\rho$,
we write the expectation value of a few-body operator $\hat{O}$ at time $t$
(the Planck constant $\hbar$ is set to unity throughout this paper):
\begin{align}\label{time_expectation}
\langle \hat{O} (t)\rangle = \Tr \left(e^{-i H t}\rho e^{i H t} \hat{O}\right) = \sum_{n,m} \rho_{nm} O_{mn} e^{-i(E_n - E_m)t} ~,
\end{align}
where $\rho_{nm}$ and $O_{mn}$ are the matrix elements of the initial state $\rho$ and the operator $\hat{O}$
in the energy eigenbasis and $E_n$ is the eigenenergy for eigenstate $n$.
Assuming no degeneracies,
off-diagonal terms dephase in the long time limit and Eq.~\eqref{time_expectation}
approaches a stationary value in the thermodynamic limit for a broad class of out-of-equilibrium initial states \cite{winter}.
The stationary value $\langle \hat{O}\rangle_{\eq}$ is the sum of diagonal elements and thus time-independent:
\begin{align}
\langle \hat{O}\rangle_{\eq} = \sum_{n} \rho_{nn} O_{nn} ~.
\end{align}
Thus, at long time the system {\it `equilibrates'}, but does not necessarily thermalize.  Examples of systems that can
`equilibrate' but not thermalize include many-body localized systems \cite{nandkishore}.

Being thermalized is a stronger statement, which means that in the thermodynamic limit the equilibrium value is equal to the average
in the corresponding thermal ensemble.  Assuming that the energy is the only extensive conserved quantity,
which fixes the temperature (and its inverse $\beta$), thermalization means the equilibrated value is
equal to the thermal (canonical (C) and microcanonical (MC)) value:
\begin{align}
\langle \hat{O}\rangle_{\eq} &=  \langle \hat{O} \rangle_{C} = \langle \hat{O}\rangle_{\MC} \label{thermalization}\\
\langle \hat{O} \rangle_{C} &= \frac{1}{Z_\beta}\Tr\left(e^{-\beta H} \hat{O}\right) \\
\langle \hat{O} \rangle_{\MC} &= \frac{1}{\mathcal{N}}\sum_{|E_n - E|\leq \Delta} O_{nn} ~,
\end{align}
where total energy $E$ and the inverse temperature $\beta$ are related by
$E = \Tr(H \rho) = (1 /Z_\beta)\Tr(e^{-\beta H} H)$ and $Z_\beta = \Tr(e^{-\beta H})$.
A macroscopically small energy $\Delta$ sets
the microcanonical energy window, and $\mathcal{N}\equiv \sum_{|E_n - E|\leq \Delta}1$
is the number of eigenstates in the window.
Equation~\eqref{thermalization}, when true, implies that {\it any} initial state with the same energy density
should thermalize the operator $\hat{O}$ to the {\it same} thermal equilibrium value.

Assuming that the above statement is true for {\it all} initial states, we may consider an extreme situation:
the initial state is an energy eigenstate ($\rho = |n\rangle\langle n|$), which is time-independent so must be thermalized.
Thus it follows that in the thermodynamic limit {\it the expectation value of every few-body operator in every energy eigenstate is the thermal value}.
This highly nontrivial statement is the essence of Eigenstate Thermalization Hypothesis (ETH) \cite{deutsch, srednicki, ro}.
Sometimes this is called `strong ETH' ({\it every} eigenstate), in contrast to `weak ETH' where some small number of eigenstates are not thermal \cite{biroli}.

One necessary condition of ETH is that the diagonal elements, $O_{nn}$, depend only on the eigenenergy
\cite{ETH_supplement}.
Therefore, after sorting the eigenstates by energy ($E_{n-1} < E_n < E_{n+1}$), we consider the following quantity:
\begin{align}\label{indicator}
r_n = \langle n+1| \hat{O} | n+1\rangle - \langle n | \hat{O} | n \rangle ~.
\end{align}
If ETH is true, this quantity should be exponentially small in the system size, because the energy difference between adjacent eigenenergies is
exponentially small.
Since the typical magnitude of $r_n$ depends on the density of states at energy $E_n$,
we consider only a `central' half of the eigenstates in the spectrum of $H$, where the variation in the density of states is small
(see Appendix A).
In section III, we examine various aspects of the parameter $r_n$, its distribution, its largest absolute values,
and features of the corresponding `outlier' eigenstates.

\subsection{Models}
We consider two nonintegrable one-dimensional Hamiltonians:

{\it Model 1.} Ising chain with transverse ($g$) and longitudinal ($h$) fields:
\begin{align}\label{ising}
H = \sum_{i=1}^L \left(g \sigma^x_i + h\sigma^z_i + J \sigma^z_i \sigma^z_{i+1}\right) ~,
\end{align}
where $\sigma^x_i$ and $\sigma^z_i$ are the Pauli matrices of the spin at site $i$.
We use periodic boundary conditions, so site $L+1 =1$.
For nonzero values of all the parameters ($g$, $h$, and $J$), this model is known to be nonintegrable;
specifically, we use the parameters $(g,h,J)$ = (0.9045, 0.8090, 1),
where this model is robustly nonintegrable for the system sizes we can exactly diagonalize \cite{hyungwon}.

As few-body operators, we look at the single-site operators $\sigma^x_i$ and $\sigma^z_i$
(the expectation value of $\sigma^y_i$ is zero due to time-reversal symmetry),
and two two-site operators, $\sigma^x_i \sigma^x_{i+1}$ and $\sigma^y_i \sigma^y_{i+1}$
(the expectation value of $\sigma^z_i \sigma^z_{i+1}$ is fixed by those of the single-site operators and the energy).
We look at momentum eigenstates, so the results do not depend on the site $i$ and we fix $i = 1$.

{\it Model 2.}  One-dimensional hard-core bosons
with nearest- and next-nearest-neighbor hopping and interaction:
\begin{align}
H = &\sum_{i=1}^L  \left[  -t(b_{i+1}^\dagger b_i + b_i^\dagger b_{i+1})  +V n_i n_{i+1} \right]\notag\\
&\quad + \sum_{i=1}^L  \left[  -t' (b_{i+2}^\dagger b_i + b_i^\dagger b_{i+2})  +V' n_i n_{i+2} \right] ~, \label{boson}
\end{align}
where $b_i$ ($b_i^\dagger$) is the annihilation (creation) operator of a hard-core boson on site $i$
with
$[b_i,b_j] = [b_i^\dagger, b_j^\dagger] = [b_i,b_j^\dagger ] = 0 \ \text{for}\ i\neq j$
and
$\{b_i,b_i\} = \{b_i^\dagger, b_i^\dagger\} = 0 \ \text{and} \  \{b_i,b_i^\dagger \} = 1$,
and $n_i \equiv b_i^\dagger b_i$ the number of bosons at site $i$.
The total number of bosons $N$ is conserved and we focus on half-filling, $N/L=1/2$ (thus only even $L$).
We impose periodic boundary conditions; L+1 = 1.
We choose $t = V = t' = V' = 1$, for which this model is known to be nonintegrable~\cite{santos2}.

As few-body operators, we look at the three two-site operators
$n_j n_{j+1}$, $(b_j^\dagger b_{j+1}+b_{j+1}^\dagger b_j)/2$,
and $(b_j^\dagger b_{j+1} - b_{j+1}^\dagger b_j )/(2i)$,
which form a basis of the observables acting on two neighboring sites whose expectation values in
simultaneous eigenstates of $H$ and momentum
are not simply dictated by this system's symmetries and conservation laws.
We only consider $j=1$, since we look only at eigenstates of momentum.
For this model, the expectation values of all of the one-site operators do not vary between these eigenstates,
and thus are not of interest for this study of ETH.

We first write each Hamiltonian in block-diagonal form in the momentum basis and diagonalize each momentum sector.
Then we collect results from all momenta and sort the exact many-body eigenstates in ascending order of energy.
Although level statistics should be carried out within each momentum sector,
ETH should be valid regardless of discrete symmetries \cite{santos}.
Since the eigenstates with positive and negative momenta map on to one another under time reversal,
we only diagonalize non-negative momenta $0\leq k\leq L/2$.  A state with momentum $k$ has eigenvalue $\exp{(i2\pi k/L)}$ under the operator that
translates the system by one lattice spacing.
The lengths of the systems we diagonalize are $L$ = 12 to 19 (18 for the Floquet operator) for Model 1,
and even $L$'s from 14 to 22 for Model 2.

\section{Results}
For the Ising model (Eq.~\eqref{ising}), we present results for the operator $\sigma^x_1$, unless otherwise specified.
For the hard-core boson model (Eq.~\eqref{boson}), we present results for the operator $n_1 n_2$, unless otherwise specified.
The results of the other three operators are qualitatively the same and given in Appendix B.

\subsection{Distribution of $|r|$}
\begin{figure}
\includegraphics[width=1.0\linewidth]{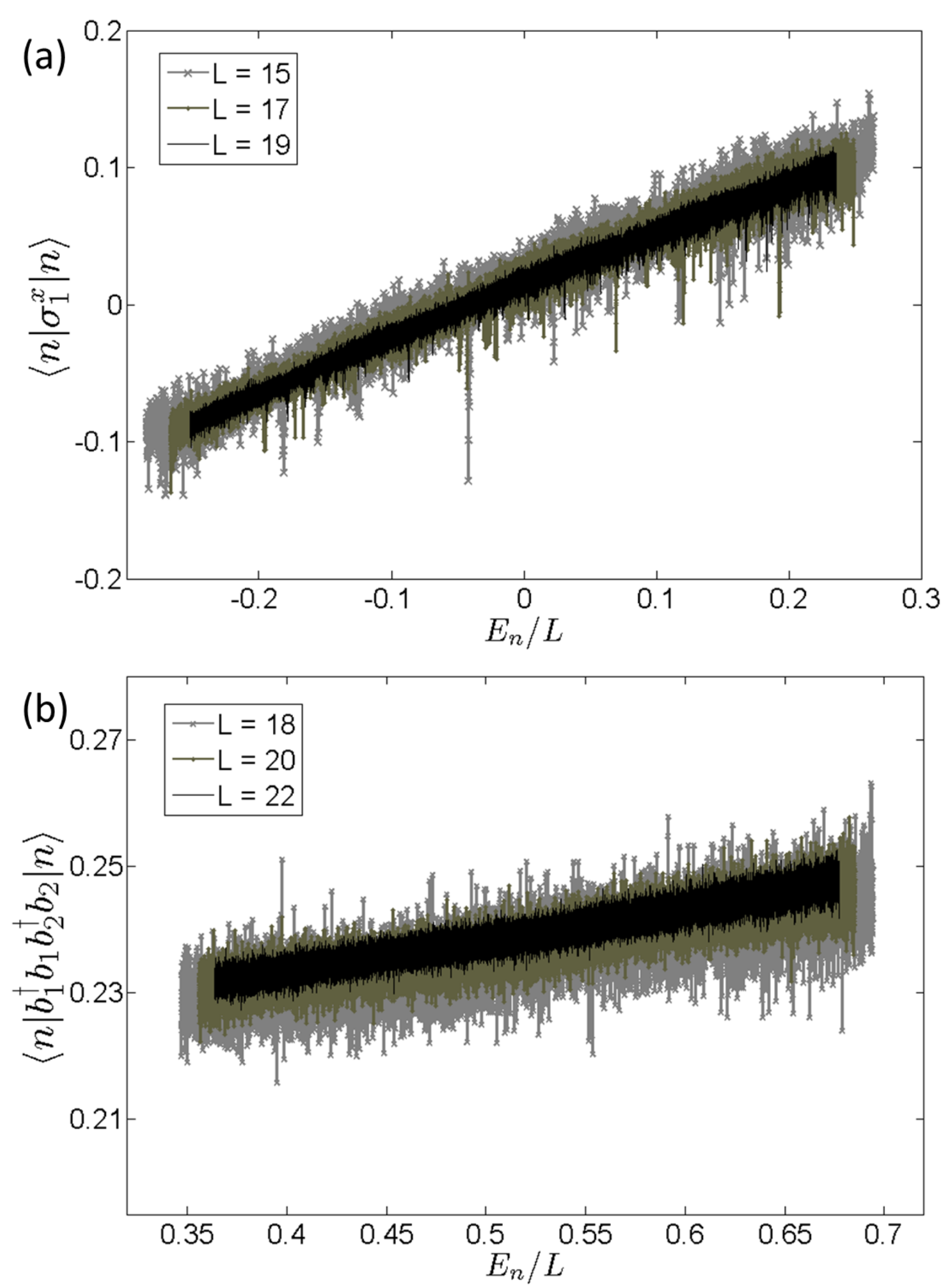}
\centering
\caption{Diagonal elements of a few-body operator in energy eigenbasis vs. energy density. The darker, the larger the system size. (a) Ising Hamiltonian (Eq.~\eqref{ising}). The operator is $\sigma^x_1$. (b) Hard-core boson Hamiltonian (Eq.~\eqref{boson}). The operator is $n_1 n_2 = b^\dag_1 b_1 b^\dag_2b_2$.
For both cases, the fluctuations become smaller as the system size increases.
}
\label{expectation}
\end{figure}
Figure \ref{expectation} shows the expectation values in each energy eigenstate vs.~its energy density.
It is clear that as we increase the system size (approaching the thermodynamic limit), these
fluctuations reduce and the expectation value becomes a smooth function of energy density.
This is in a good agreement with ETH predictions.
Note that Refs.~\cite{ro,rigol1} shows the fluctuations are not small in an integrable system.

\begin{figure}
\includegraphics[width=1.0\linewidth]{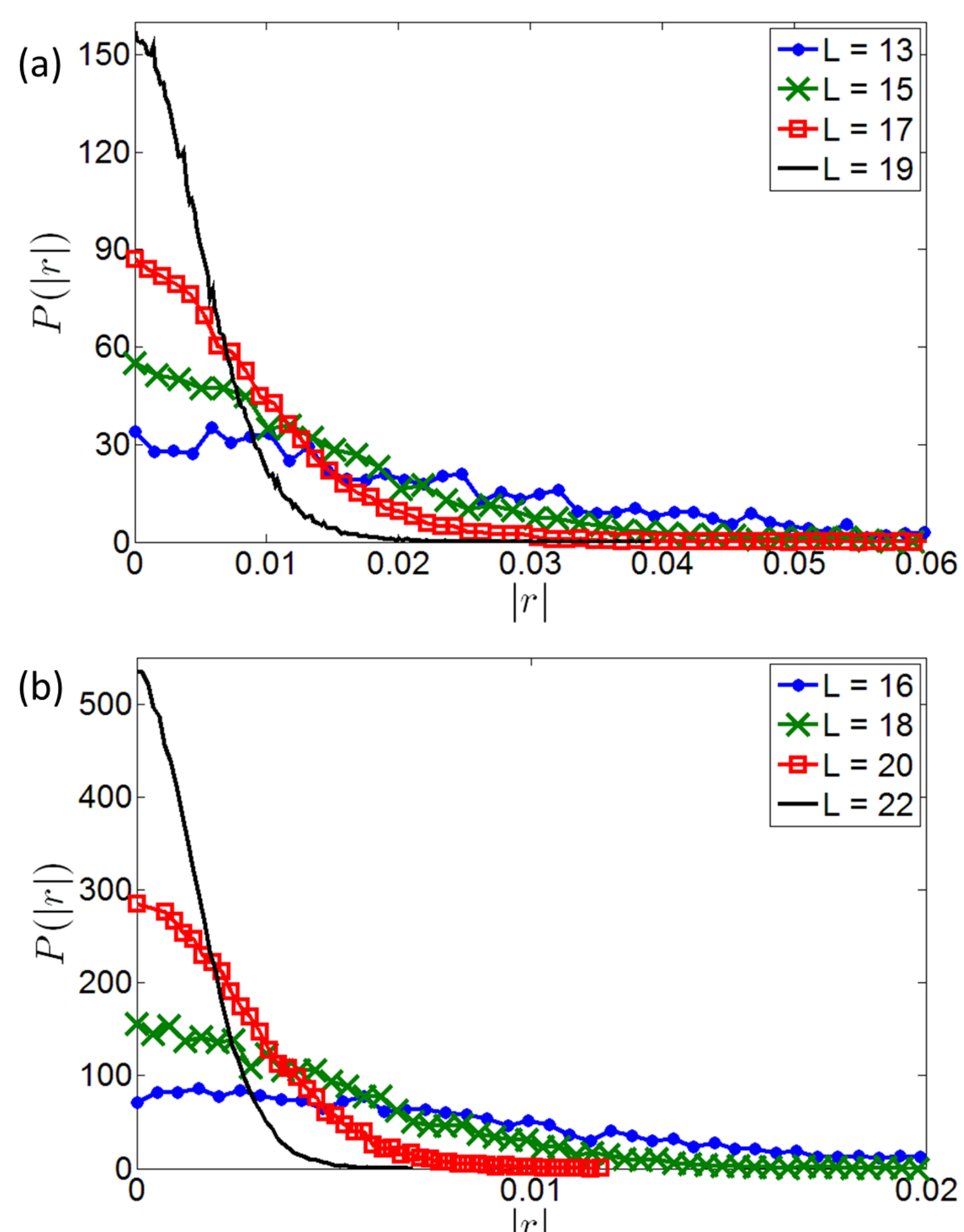}
\centering
\caption{(color online) Distribution of $|r|$. (a) Ising Hamiltonian (Eq.~\eqref{ising}), and (b) Hard-core boson Hamiltonian (Eq.~\eqref{boson}).
As we increase the system size, the distribution becomes sharply peaked near $|r| = 0$.
The distribution can be well-fitted by a Gaussian distribution (only positive argument) with a standard deviation $\sigma$ decreasing exponentially with the system size $L$.}
\label{distribution}
\end{figure}

Next, we compute the distribution of $|r|$ (state index $n$ is omitted when the meaning is straightforward)
and see how it behaves as we increase the system size.
Figure \ref{distribution} is the plot of the distributions of $|r|$.
We see that the distribution becomes sharply peaked at $|r| = 0$ for a larger system size,
which is consistent with ETH.

To quantify the fluctuations, we consider the average of $|r_n|$
(the average of $r_n$ is basically zero).
Note that $r_n$ is an indicator of ETH without microcanonical averaging,
thus it can be considered as an extreme version of the indicator introduced in Refs. \cite{santos, khatami1, beugeling}
that is averaged over a small energy window.
As we can see in Figure \ref{outliers}, the mean value of $|r|$
decreases exponentially with the system size (thus a power-law decay with the Hilbert space dimension), in accordance with ETH \cite{fitting}.

As long as the thermodynamic limit of the distribution $P(r)$ is the Dirac delta function $\delta(r)$,
the conventional microcanonical formulation of equilibrium statistical mechanics is valid, since
averaging a few-body operator over a microcanonical ensemble,
which still includes exponentially many states even in a narrow energy window,
should be equal to the canonical ensemble average.
Therefore, Figure \ref{distribution} provides numerical evidence of the validity of
equilibrium statistical mechanics for these isolated quantum systems.
This feature is also confirmed by several previous works \cite{ro, santos, beugeling}.
This implies that {\it almost all} out-of-equilibrium initial states
will eventually equilibrate and thermalize in terms of the expectation values of few-body operators.
This is enough for all practical purposes, since it is impossible to precisely manipulate
highly-excited many-body states.
Sometimes, this is called ``weak ETH''.

Here we ask whether there could be some small number of eigenstates with
nonzero finite values of $r_n$
in the thermodynamic limit.
Such `outlier' states if sufficiently rare would not contribute to the microcanonical ensemble average,
and thus not compromise standard equilibrium statistical mechanics.
However, Eigenstate Thermalization Hypothesis is stricter:
{\it every} eigenstate is thermal.  This is sometimes referred as ``strong ETH''.
Strong ETH requires that {\it every} $r_n$ approaches zero in the thermodynamic limit.

\subsection{Outliers of $|r|$}
\begin{figure}
\includegraphics[width=1.0\linewidth]{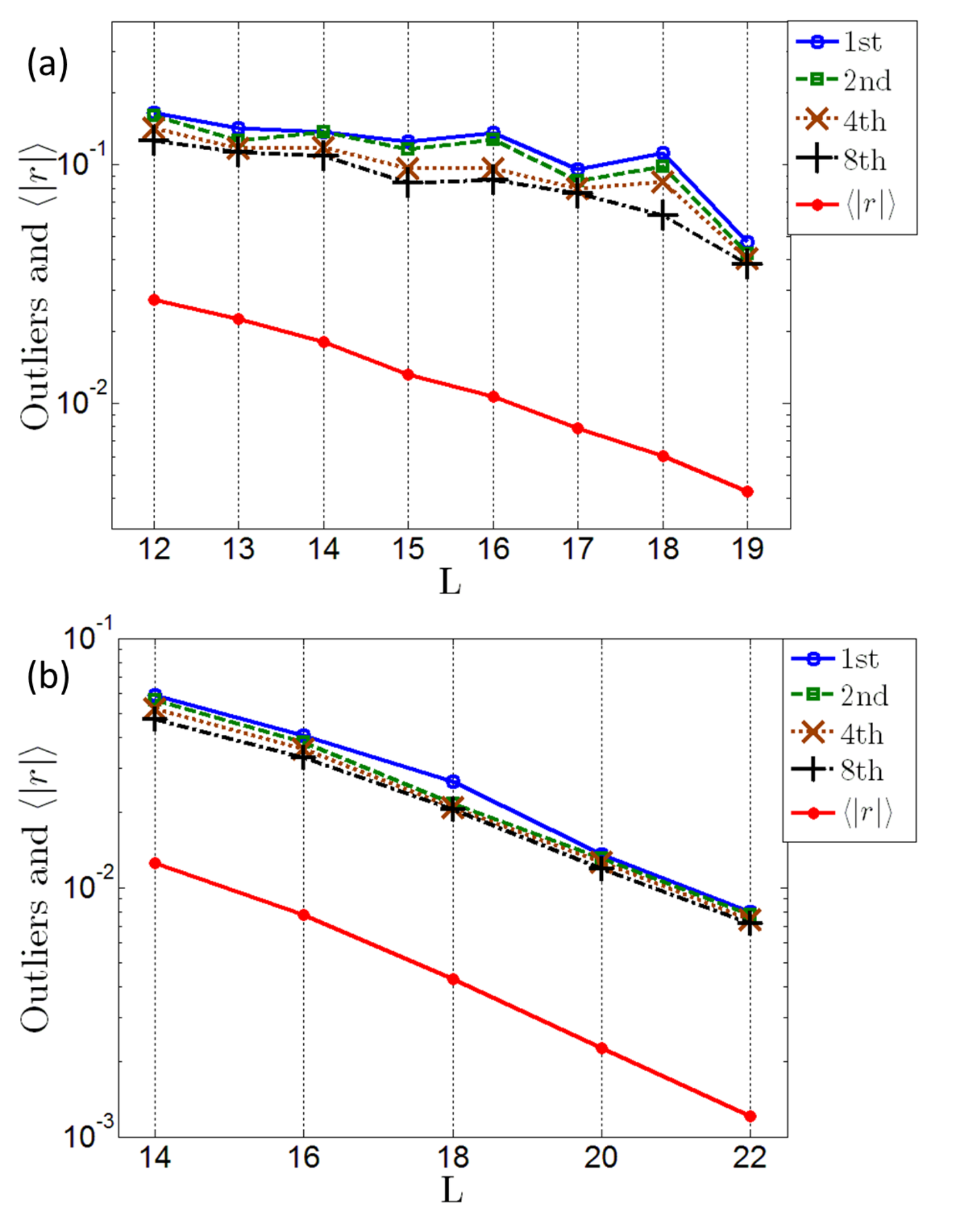}
\centering
\caption{(color online) From top to bottom in each figure: 1st, 2nd, 4th, 8th largest value of $|r|$ and the mean value $\langle |r| \rangle$. (a) Ising Hamiltonian (Eq.~\eqref{ising}),
and (b) Hard-core boson (Eq.~\eqref{boson}).
The mean value $\langle |r| \rangle$ decreases exponentially in $L$ as ETH suggests \cite{beugeling}.
The largest `outliers' also decrease, although slower than $\langle |r| \rangle$, with the system size.
}
\label{outliers}
\end{figure}

We introduce one way to numerically test the strong ETH: the outliers.
We define the outliers as the eigenstates that deviate the most from ETH and thus give large values of $|r_n|$.
Since each $r_n$ comes from two states,
a large $|r_n|$ means either one state is an outlier and the other one is ``normal'' or
both states are outliers of opposite signs.
This also implies that the indicator $|r_n|$ does not detect the special case where
two outliers of the same sign are consecutive in the energy spectrum, resulting in a small $|r_n|$.
In such a case, however, both $|r_{n-1}|$ and $|r_{n+1}|$ should be large so we can easily identify the outliers \cite{worst_case}.
Usually, we find one outlier state next to many normal states. Therefore, there are no difficulties in
spotting these outlier states.
If the strong ETH is correct,
even the largest value of $|r_n|$ should decrease to zero in the thermodynamic limit.

Figure \ref{outliers} shows the 1st, 2nd, 4th, and 8th largest values of $|r_n|$ and
the mean value of $|r_n|$. It is clear that the mean value $\langle |r| \rangle$
(averaged over half of the spectrum) decreases exponentially with the system size.
This shows that our (isolated) quantum models agree well with the microcanonical formulation of equilibrium statistical mechanics,
as explained earlier.
The central feature of Figure \ref{outliers}
is that even the largest value of $|r|$ decreases with the system size.
Extrapolating this tendency, we provide supporting evidence for the strong ETH, that
every eigenstate far away from the edge of the spectrum is thermal as far as the expectation values of few-body operators are concerned.

Now let us examine these outlier states.
Table \ref{outlier_table_ising} is the list of the 1st, 2nd, 4th, and 8th outlier states and their properties of
four few-body operators for the Ising Hamiltonian (Eq.~\eqref{ising}) with $L$ = 18.
We consider three features; (1) the location of the outlier state in the list of the states considered
sorted by energy in ascending order (state $\#$),
(2) participation ratio (PR), (3) and momentum $k$.

(1): The state number tells us where the outliers are located in the spectrum.
It shows that the outlier states may come from any place in the spectrum,
which implies that their presence is not strongly sensitive to the small changes in the density of states over the energy range we are studying.
Another feature is that different operators do share outlier states:
States $\#$ 70164 and 8033 are the two most extreme outlier states for all of $\sigma^x_1$, $\sigma^z_1$ and $\sigma^x_1\sigma^x_2$,
but are not outlier states for $\sigma^y_1\sigma^y_2$.
We have checked the 30 most extreme outlier states for $L$ = 18 and it turns out that
$\sigma^x_1$, $\sigma^z_1$, $\sigma^x_1\sigma^x_2$ share many outlier states, while
$\sigma^y_1 \sigma^y_2$ seldom has common outlier states with the other operators.

\begin{table}
\begin{tabular}{cccccc}
\hline
$L$ = 18 & &  & & & $k:0 \sim 9$ \\
\hline
 & & state \# & & PR ($\times 10^4$)  & $k$  \\
\hline
${\hat O}$ = $\sigma^x_1$ & & & & \\
1st & & 70164 & & 0.907 & 9 \\
2nd & & 8033 &  & 1.17 & 0 \\
4th & & 48351 &  & 1.37 & 9 \\
8th & & 69512&  & 7.33 & 6 \\
\hline
${\hat O}$ = $\sigma^z_1$ & & & & & \\
1st & & 8033 &  & 1.17 & 0 \\
2nd & & 70164 &  & 0.907 & 9 \\
4th & & 52061 &  & 1.72 & 9 \\
8th & & 64591&  & 4.44 & 0 \\
\hline
${\hat O}$ = $\sigma^x_1\sigma^x_2$ & & & & & \\
1st & & 8033 &  & 1.17 & 0 \\
2nd & & 70164 &  & 0.907 & 9 \\
4th & & 48351 &  & 1.37 & 9 \\
8th & & 8014&  & 2.89 & 4 \\
\hline
${\hat O}$ = $\sigma^y_1\sigma^y_2$ & & & & & \\
1st & & 64675 &  & 4.82 & 0 \\
2nd & & 70607 &  & 4.86 & 0 \\
4th & & 72440 &  & 4.73 & 0 \\
8th & & 20383&  & 5.66 & 9 \\
\hline
\end{tabular}
\caption{Ising Hamiltonian (Eq.~\eqref{ising}).
List of 1st, 2nd, 4th and 8th outliers for four few-body operators and their state $\#$,
participation ratio (PR), and the momentum of the state ($k$). The state $\#$ is the location of the state in the list of the analyzed states sorted by ascending order in energy. For $L$ = 18, total 72821 states are considered. }
\label{outlier_table_ising}
\end{table}

\begin{table}
\begin{tabular}{cccccc}
\hline
$L$ = 20 & &  & & & $k:0 \sim 10$ \\
\hline
 & & state \# & & PR ($\times 10^4$)  & $k$  \\
\hline
${\hat O}$ = $b_1^\dagger b_1 b_2^\dagger b_2$ & & & & & \\
1st && 48359 && 5.29 &  0\\
2nd && 50735 && 5.49 & 0\\
4th && 49503 && 8.55 & 9\\
8th && 48585 && 7.90 & 3\\
\hline
${\hat O}$ = $(b_1^\dagger b_2 +b_2^\dagger b_1)/2$ & & & & \\
1st && 48993 && 5.17 & 0\\
2nd && 45849 && 5.77 & 0\\
4th && 47993 && 5.60 & 0\\
8th && 48766 && 5.03 & 0\\
\hline
${\hat O}$ = $(b_1^\dagger b_2 -b_2^\dagger b_1)/(2i)$ & & & & & \\
1st && 35990 && 8.70 & 5\\
2nd && 42699 && 8.65 & 9\\
4th && 34191 && 8.39 & 3 \\
8th && 31557 && 8.55 & 7\\
\hline
\end{tabular}
\caption{Boson Hamiltonian (Eq.~\eqref{boson}).
List of 1st, 2nd, 4th and 8th outliers for three few-body operators and their state $\#$,
participation ratio (PR), and the momentum of the state ($k$). The state $\#$ is the location of the state in the list of the analyzed states sorted by ascending order in energy. For $L$ = 20, total 50814 states are considered.
The operator $(b^\dag_1 b_2 - b_2^\dag b_1)/(2i)$ is odd under time reversal thus has zero expectation value for $k = 0$ and $10$ states,
which are even under time reversal. }
\label{outlier_table_boson}
\end{table}

(2) and (3): The participation ratio (PR) quantifies how delocalized a state is in a certain basis \cite{wegner}.
Here we write each eigenstate in the $\sigma^z_i$ basis for the Ising model,
and in the $n_i$ basis for the boson model.
Then, for a given eigenstate $|n\rangle = \sum_{s=1}^D c_s^{(n)} |s\rangle$ ($D$ is the Hilbert space dimension, $|s\rangle$ are the basis states), the PR is
\begin{align}
{\rm PR} = \frac{1}{\sum_{s = 1}^D |c_s^{(n)}|^4} ~.
\end{align}
When the state is completely delocalized, each $|c_s^{(n)}|^2 = 1/D$ and thus PR = $D$.
If the state is totally localized on only one basis state, then PR = 1.  Thus the PR is a measure of how many basis states the state is delocalized over.

There are two special momenta, $k = 0$ and $k = L/2$, which preserve the system's time-reversal symmetry (for odd $L$, only $k=0$ exists).
These two momentum sectors have an additional discrete symmetry, spatial inversion,
and thus each eigenstate in these sectors is either even or odd under this symmetry.
Consequently, this extra discrete symmetry prevents these special momentum eigenstates from
exploring all basis states, and results in a smaller PR than the states with the other momenta.
For $L$ = 18,
the average PR of the central half spectrum with momenta $k = 1 \sim 8$ is 8.53$\times 10^{4}$ with standard deviation 0.59$\times 10^4$,
while the average PR of states with momenta $k = 0$ and $k=9$ is 5.55$\times 10^{4}$ with standard deviation 0.45$\times 10^4$ \cite{average_spectrum}.
Therefore, these special momentum states are less ergodic and are good candidates for outliers.
As we can see in Table \ref{outlier_table_ising}, many (but not all) outliers do have small PR
compared to the average value of PR for their momenta,
which implies these states are less uniformly delocalized than typical states.
Also, outlier states tend to come from the special time-reversal-symmetric momenta, as expected.
Nevertheless, Figure \ref{outliers} indicates that
even these extreme outlier states seem to obey ETH in the thermodynamic limit.
The tendencies of outlier states are similar for other $L$'s.

Table \ref{outlier_table_boson} is the list of outlier states and their PR and momenta for the boson Hamiltonian (Eq.~\eqref{boson}) with $L = 20$.
The average PR for $k = 0$ and $L/2 = 10$ is $5.50\times 10^4$ with standard deviation $0.30\times 10^4$,
and the average for the other $k$'s is $8.61\times 10^4$ with standard deviation $0.27 \times 10^4$.
Again, many of the outlier states are from the special momenta ($k = 0$ and $L/2$).
But in contrast to the Ising model, the PR's of extreme outliers are not much lower than the averages.
This indicates that the PR need not be extreme in outlier states
\cite{entanglement}.
It turns out that the lowest PR state for the boson Hamiltonian
($\mathrm{PR} = 1.19\times 10^4$ for $L$ = 20) is not extreme in the $|r|$ measure for these few-body operators,
while the lowest PR state in the Ising Hamiltonian (state $\#$ = 70164 in Table \ref{outlier_table_ising})
is an outlier in $|r|$ for most of the few-body operators we examined.
Note that the local current operator $(b^\dag_1 b_2 - b^\dag_2 b_1)/(2i)$ is odd under time-reversal symmetry
and thus has zero expectation value for all states that are symmetric under time reversal.
Therefore, $k = 0$ and $L/2$ states cannot have outliers of $(b^\dag_1 b_2 - b^\dag_2 b_1)/(2i)$.
Since these special momentum states are not outliers for this operator, it has smaller value of extreme $|r|$'s
(see Appendix B for extreme values).

In Figure \ref{outliers}(a), there is an alternation between
even and odd $L$ in outlier values of $|r|$.
We attribute this to the number of special momenta.  Even chains have two special momenta ($k = 0$ and $L/2$),
while odd chains have only $k=0$.  As we see from Table \ref{outlier_table_ising},
these special momentum states tend to have more extreme outliers.
Since even chains have more such states, it is natural that they tend to have higher extreme values of $|r|$.
Nevertheless, once we decompose the data by the parity of $L$, we see that the largest value of $|r|$ decreases monotonically with $L$
for both even and odd $L$.

\section{Role of energy conservation in ETH: Floquet system}
Few-body conservation laws put constraints on a system's dynamics that can slow down or impede thermalization.  One example is integrable systems that have many such conservation laws and thus do not fully thermalize \cite{rigol, calabrese}.
Our Ising model has only one few-body conservation law, which is the energy (its Hamiltonian is a sum of one- and two-body operators).
Thus, we can ask what happens to thermalization when we remove energy conservation, so that the system has no few-body conserved quantity.
[Note that these systems always have all the many-body conserved operators that are the projections on to the system's eigenstates.]

One way to remove energy conservation but still have eigenstates of the dynamics is to have a time-dependent Hamiltonian $H(t)$ that is periodic in time, so $H(t)=H(t+\tau)$.
We can divide the Ising Hamiltonian (Eq.~\eqref{ising}) into
two parts, one with only $\sigma^x$ operators ($H_x = \sum_i g \sigma^x_i$) and one with only $\sigma^z$ operators ($H_z = \sum_i h\sigma^z_i  + \sigma^z_i\sigma^z_{i+1}$).
During one period $\tau$, we have $H(t)=H_z$ for the first $\tau/2$ and then $H(t)=H_x$ for the remaining $\tau/2$.
Then, the unitary Floquet operator that takes the system through one period is
\begin{align}
\hat{U} = \exp\left(- i H_x \tau/2\right)\exp\left(-i H_z \tau/2\right) ~.
\end{align}

The eigenvalues of $\hat{U}$ are complex numbers of magnitude one.
On the complex unit circle, the level spacing statistics of the eigenvalues follow the `circular orthogonal ensemble' \cite{luca}.
References \cite{lazarides,ponte} report that when this sort of many-body Floquet system thermalizes,
it thermalizes a generic initial state to the `infinite temperature' ensemble
(all states of each small subsystem equally probable).
Therefore, if the strong ETH is applicable to this Floquet system,
every eigenstate of $\hat{U}$ should be thermal at infinite temperature.

Here we perform the same outlier analysis for eigenstates of $\hat{U}$ with the few-body operator $\sigma^x_1$.
Since the expectation value of $\sigma^x_1$ at infinite temperature is zero,
we need not compare expectation values of adjacent states and instead
just evaluate the absolute value of the expectation value in each eigenstate.
Furthermore, since the eigenvalues are well spread over the unit circle,
we can use all eigenstates and need not search for a region where the density of states is roughly constant.
We choose the period $\tau = 1.6$ and use the same parameters $g,h,J$ as above.

\begin{figure}
\includegraphics[width=1.0\linewidth]{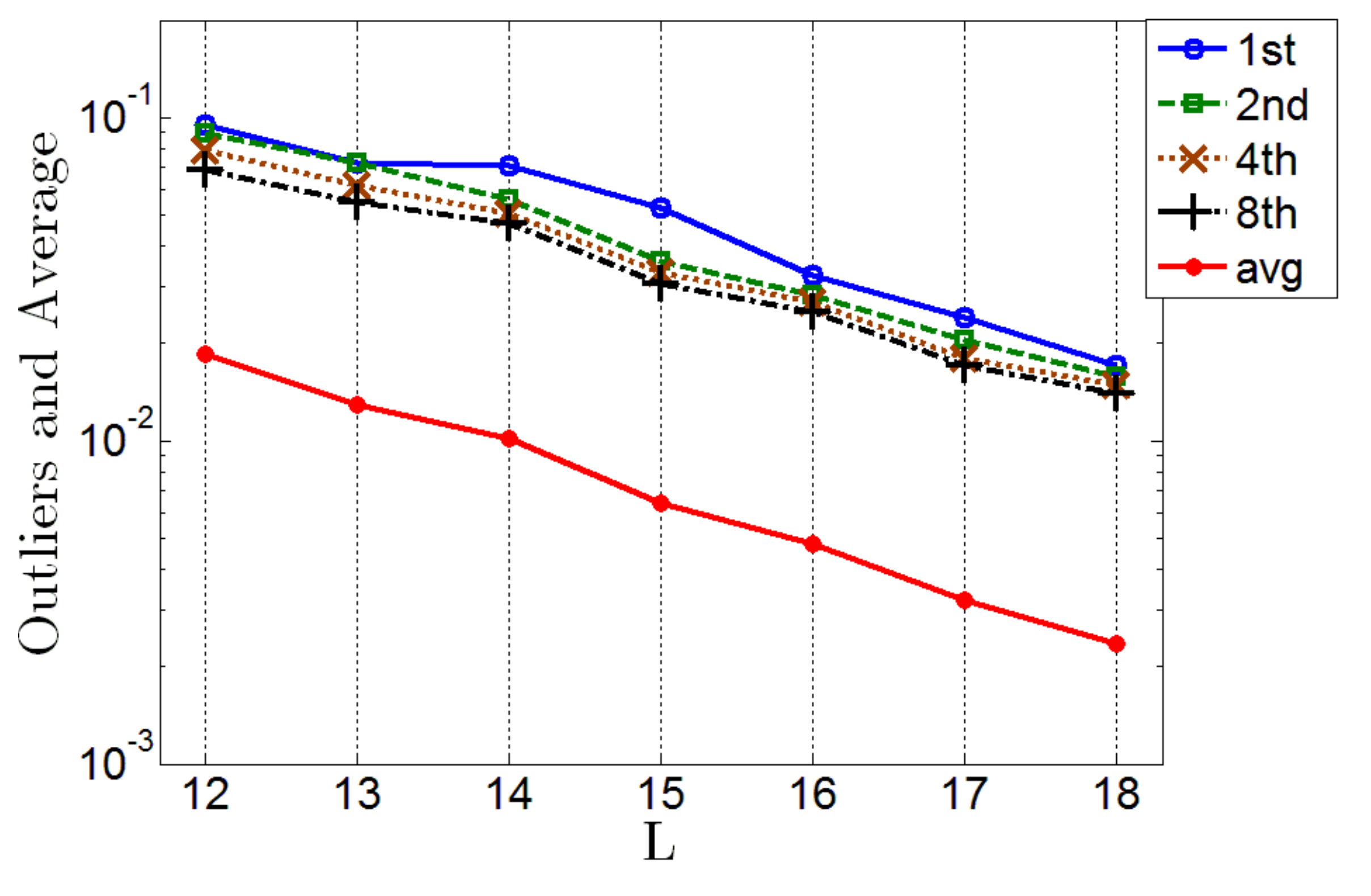}
\centering
\caption{(color online) From top to bottom: 1st, 2nd, 4th, and 8th outliers and the average of absolute value of eigenstate expectation value for the Floquet operator $\hat{U}$.
The few-body operator is $\sigma^x_1$. The average value and the outlier values are substantially smaller than those of the corresponding
Hamiltonian.
Therefore, the Floquet system satisfies ETH more precisely for each system size $L$. }
\label{floquet}
\end{figure}

Figure \ref{floquet} shows the outliers and average value of the magnitude of the eigenstate expectation value $|\langle\sigma^x_1\rangle|$.
In comparison with Figure \ref{outliers}(a), we can clearly see that the value of outliers and average value
are smaller and approach ETH predictions faster.
Therefore, we conclude that the Floquet system, which has no local conservation law,
thermalizes ``better'' (by this measure) than the nonintegrable Hamiltonian system with energy conservation.

\section{Conclusion and Outlook}
In this paper, we have done a stringent test of the Eigenstate Thermalization Hypothesis.
We chose two popular nonintegrable model Hamiltonians, well away from integrable points in their parameter spaces,
and thoroughly investigated properties of eigenstate expectation values of several few-body operators.
First, we introduced an indicator that measures deviation from ETH behavior,
and showed that these deviations
decrease as we increase the system size.  Therefore, we recover some known results about ETH \cite{ro, beugeling}.
Then, we examined the `outlier' eigenstates, that
deviate the most from ETH behavior.
Even these extreme states (outliers) approach ETH behavior as the system size is increased.
Thus we provide numerical evidence in support of ETH in its strong version: ETH is true for {\it all} eigenstates.
We analyzed the outlier states and showed outliers have relatively small
participation ratio, which implies they are less delocalized in the many-body Fock space than typical states.
Finally, we deliberately broke the energy conservation by making the Hamiltonian time-dependent (Floquet system).
We showed that the eigenstates of the Floquet operator deviate from ETH behavior by less than those of the corresponding time-independent Hamiltonian.
Therefore, the Floquet system, which has no conserved energy to transport, thermalizes better.

Many open questions still remain. First, we have only considered certain one-site and two-site operators.
One could in principle search systematically over some complete set of few-body operators to find the
operators that produce the most extreme outliers.
We have not done this, so the possibility remains that ETH might fail for the combination of special eigenstates and special few-body operators, although we see no reason to expect such a failure.
Second, we only looked at two models, and there might be some other nonintegrable models that would violate the strong ETH.
We know that strong ETH is false for integrable models, while it
appears to be true for the two strongly nonintegrable models we have studied.
Naively, we would expect that ETH is restored in the thermodynamic limit as soon as the integrability is broken by some nonzero amount,
but this remains an open question
\cite{rigol1, neuenhahn}.
We leave these interesting questions for future investigation.

\section{Acknowledgement}

We thank Alexey Gorshkov, Vadim Oganesyan, Marcos Rigol, and David Weiss for stimulating discussions and suggestions.
H.K. thanks Samsung Scholarship for financial support. T.N.I. acknowledges the JSPS for financial support (Grant No. 248408).

\appendix
\begin{figure}
\includegraphics[width=1.0\linewidth]{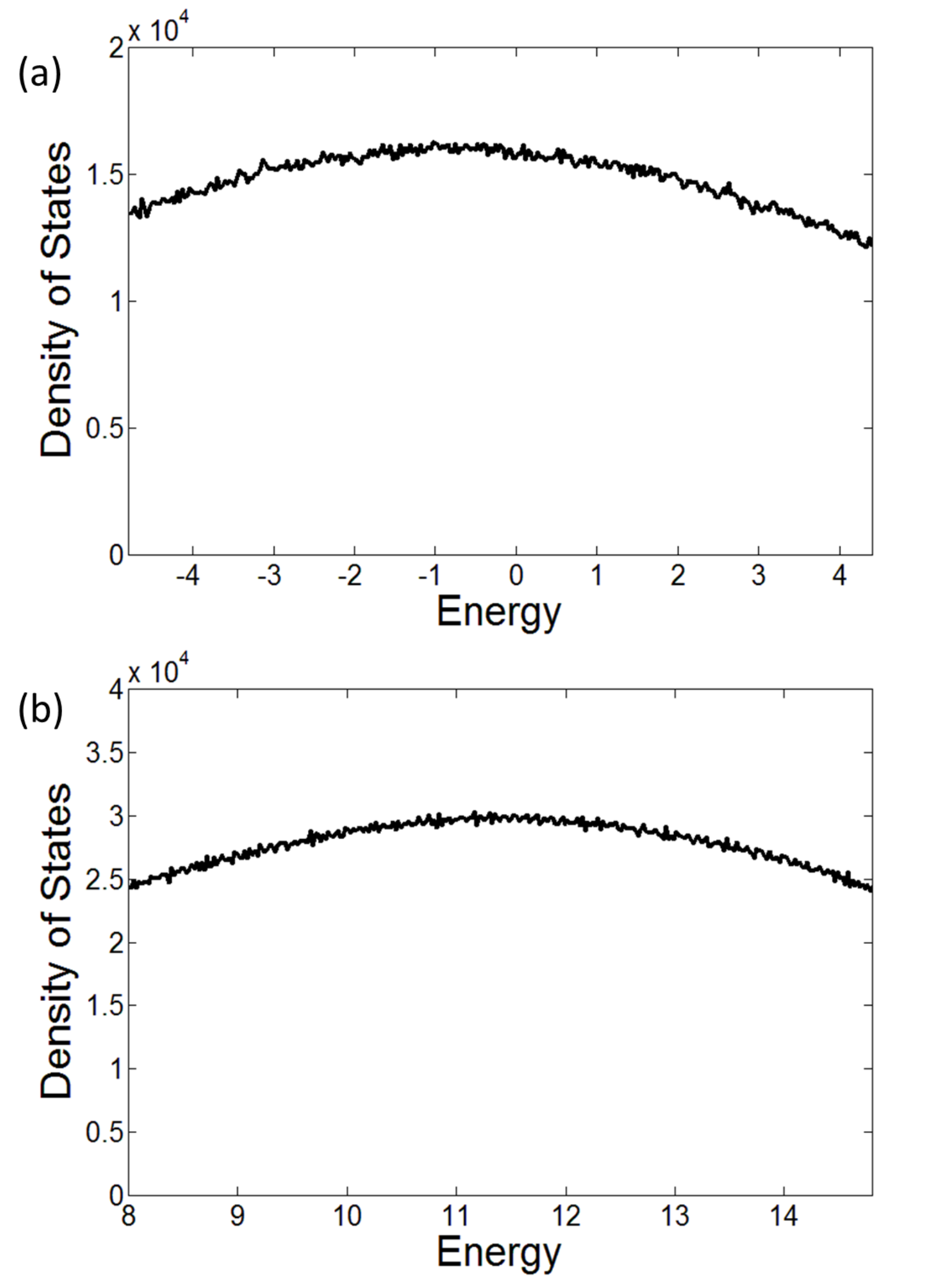}
\centering
\caption{The many-body density of states over the energy ranges we examine.  (a) Ising chain (Eq.~\eqref{ising}) with $L = 19$.
(b) Hard-core bosons (Eq.~\eqref{boson}) with $L = 22$.
For both Hamiltonians the ratio of maximum to minimum density of states over our energy range is $\sim$ 1.20. }
\label{doe}
\end{figure}

\section{Density of states and ETH indicator}

In Ref. \cite{beugeling}, they report that the typical deviations of eigenstate expectation values from their thermal values
scale as $D^{-1/2}$, where $D$ is the dimension of the Hilbert space.  For the Hamiltonian systems with thermal eigenstates that are
in a certain sense microcanonical, $D$
is proportional to $e^{S(E)}$ and thus to the many-body density of states, where $S(E)$ is the thermodynamic entropy at energy $E$.
Therefore, we expect that typical values of our ETH indicator, $|r_n|$, are inversely proportional to the square root of the density of states.
The density of states does vary over the energy range we study and thus
direct comparison of `bare' values of $|r_n|$ could be dangerous,
especially near the edge of the spectrum where the variation in the density of states is the largest.
This is why we choose to only look at one half of the eigenstates, and only those that are near the middle of the spectrum where the density of states is nearly constant.  The states we left out are the lowest and highest 25\% of the eigenstates for the Ising model, while for the boson model, which has the maximum in its density of states more substantially off center, we left out the lowest 30\% and the highest 20\%.
Figure \ref{doe} shows the densities of states over the energy range we kept.
It shows that the variations in densities of states are small in this range; the fractional difference between the largest and the smallest is less than 20$\%$.  There is a small resulting tendency for the outliers to be more likely where the density of states is lower, but because the density of states is nearly constant over the range we examine, this is a small effect.

\section{Results of other observables}
\begin{figure}
\includegraphics[width=1.0\linewidth]{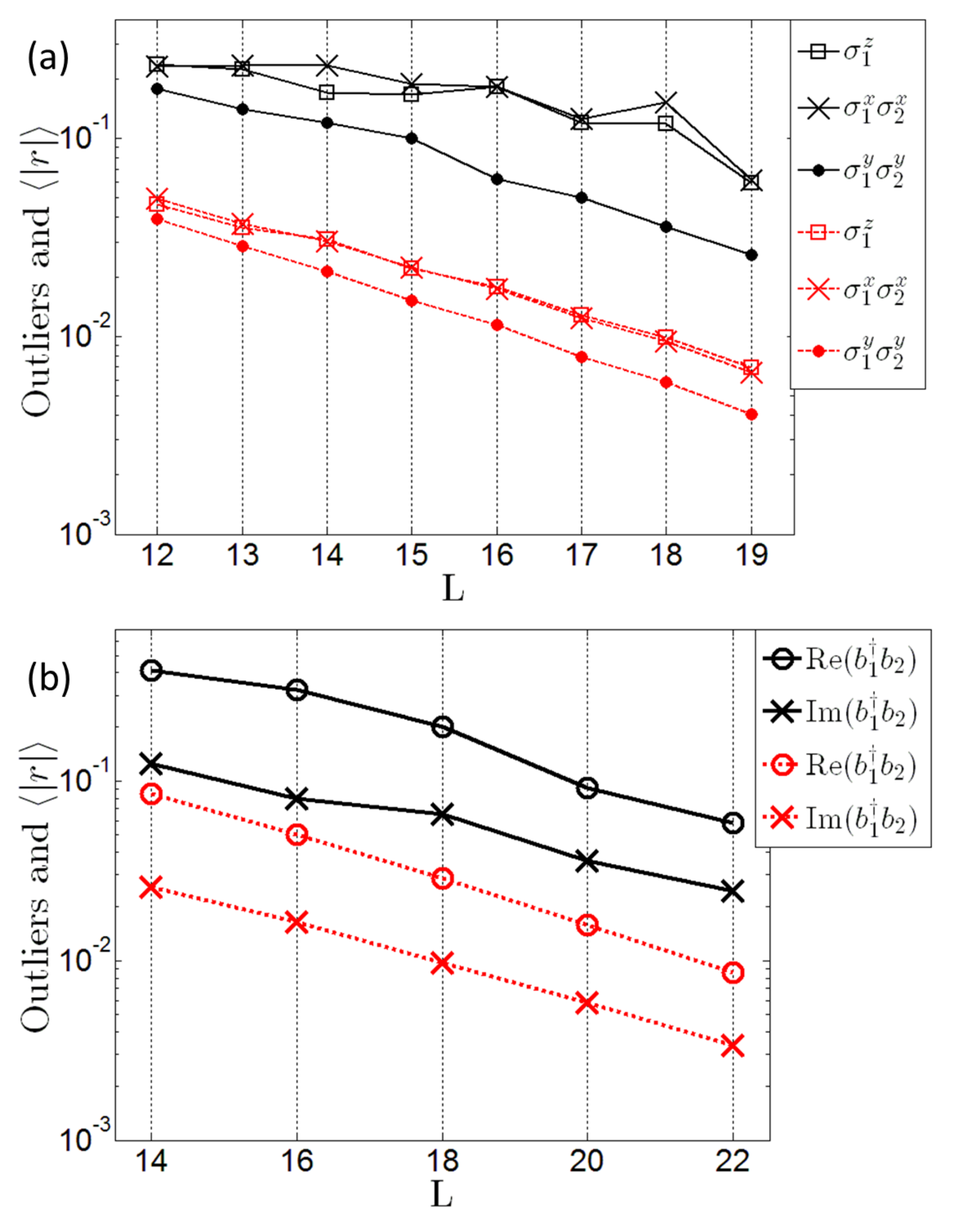}
\centering
\caption{(color online) The largest outlier values (black solid lines) and the mean of $|r|$ (red dotted lines) for other few-body operators. (a) Ising chain (Eq.~\eqref{ising}).  Circles, squares and diamonds are results for $\sigma^z_1$, $\sigma^x_1\sigma^x_2$ and $\sigma^y_1\sigma^y_2$, respectively.
(b) Hard-core bosons (Eq.~\eqref{boson}).  Circles and squares are results for the real and imaginary parts of $b^\dag_1 b_2$, respectively.  For all cases, the largest outliers decrease with the system size, thus again supporting the strong ETH. }
\label{other_outliers}
\end{figure}

In the main text, we reported results for the local operators $\sigma^x_1$ for the Ising model and $b^\dag_{1}b_1 b^\dag_{2}b_{2}$ for the hard-core boson model.
Here, we give outlier results for other few body operators; $\sigma^z_1$, $\sigma^x_1\sigma^x_2$, and $\sigma^y_1\sigma^y_2$ for the Ising model
and $(b^\dag_1 b_2 + b^\dag_2 b_1)/2$ = Re($b^\dag_1 b_2$) and $(b^\dag_1 b_2 - b^\dag_2 b_1)/(2i)$ = Im($b^\dag_1 b_2$) for the hard-core boson model.
Figure \ref{other_outliers} shows the largest value of $|r|$ (the extreme outlier) and the mean value of $|r|$ for each operator.
It is clear that they have the same qualitative feature, the decrease with increasing system size of the value for the outliers.
Therefore, all results support the strong ETH.


\begin{thebibliography}{99}
\bibitem{thermal_states}
For some models that are nonintegrable and thermal in the bulk of their many-body spectrum, there may be low-energy states just above
the ground state (and just below the highest-energy state) that are asymptotically integrable and fail to obey the ETH even in the
thermodynamic limit.  Particularly, in many one-dimensional systems the low-energy dynamics are known to be asymptotically integrable.
Thus when we say we are testing whether all eigenstates are thermal, we mean in the bulk of the many-body spectrum, excluding these possible
special states near the edge of the spectrum.
\bibitem{ETH_supplement}
ETH contains another statement that the off-diagonal elements, $O_{n,m}$ ($n\neq m$), are exponentially small in
the Hilbert space dimension $D$ \cite{deutsch,srednicki,ro,khatami2}, which makes their contribution to the long-time expectation value vanish
for $D \rightarrow\infty$.
However, Ref. \cite{winter} has proven equilibration without this assumption and we do not consider this aspect here.
\bibitem{fitting}
Since our main focus is on the outliers, as is detailed in the following subection III.B,
we do not attempt to perform a precise fitting to estimate the exponent of power-law decay.
For the finite-size scaling of a closely-related measure of the deviations from ETH, see Ref. \cite{beugeling}.
\bibitem{worst_case}
There could be a consecutive sequence of outliers but the probability of this is so low that it does not occur in our data.
\bibitem{average_spectrum}
The variation of the average value of the PR with energy is smaller than these standard deviations.
The cited values in the main text are averages over the full energy range that we study.
\bibitem{entanglement}
We have also examined the bi-partite entanglement entropy of these outliers and they are also close to typical values.
\end{thebibliography}
\end{document}